\documentclass[aps,pre,twocolumn,superscriptaddress,showpacs,floatfix]{revtex4-2}
\usepackage{xcolor}
\usepackage{amssymb}
\usepackage{amsmath}
\usepackage{bm}
\usepackage{url}
\usepackage{graphicx}

\begin{document}
	
\title{  Non-equilibrium statistical mechanics of the turbulent energy cascade: \\ irreversibility and response functions}
	
\author{Niccol\`o  Cocciaglia}
\affiliation{Dipartimento di Fisica,  Universit\`a degli Studi di Roma ``Sapienza'', P.le Aldo Moro 5, 00185, Rome, Italy}

\author{Massimo Cencini}\thanks{Corresponding author: massimo.cencini@cnr.it}
\affiliation{Istituto  dei  Sistemi  Complessi,  CNR,  Via  dei  Taurini  19,  00185  Rome,  Italy}
\affiliation{INFN ``Tor Vergata'' Via della Ricerca Scientifica 1, 00133 Rome, Italy}

\author{Angelo Vulpiani}
\affiliation{Dipartimento di `Fisica,  Universit\`a degli Studi di Roma ``Sapienza'', P.le Aldo Moro 5, 00185, Rome, Italy}

\date{\today}
	
\begin{abstract}
The statistical properties of turbulent flows are fundamentally
different from those of systems at equilibrium due to the presence of
an energy flux from the scales of injection to those where energy is
dissipated by the viscous forces: a scenario dubbed ``direct energy
cascade''.  From a statistical mechanics point of view, the cascade
picture prevents the existence of detailed balance, which holds at
equilibrium, e.g. in the inviscid and unforced case.  Here, we aim at
characterizing the non-equilibrium properties of turbulent cascades in
a shell model of turbulence by studying an asymmetric
time-correlation function and the relaxation behavior of an energy
perturbation, measured at scales smaller or larger than the perturbed
one.  We shall contrast the behavior of these two observables in both non-equilibrium (forced and dissipated) and
equilibrium (inviscid and unforced) cases. Finally, we shall show that
equilibrium and non-equilibrium physics coexist in the same system,
namely at scales larger and smaller, respectively, of the forcing
scale.
\end{abstract}
\maketitle

\section{Introduction\label{Introduction}}

Understanding non-equilibrium systems is one of the most challenging
open problem of modern statistical mechanics
\cite{kubo2012statistical,livi2017nonequilibrium}.  A key aspect of
non equilibrium phenomena is the presence of currents induced by some
external constraints, which entail the breaking of detailed balance
and, consequently, of the time-reversal symmetry, or, equivalently,
imply the positivity of entropy production
\cite{ruelle1996positivity,seifert2012stochastic,peliti2021stochastic}. Measuring entropy production is not an
easy task as it requires to measure the log-ratio between the
probability of a long trajectory of the system and that of its time
reversed \cite{lebowitz1999gallavotti}. This is feasible in relatively
simple Markov models but it is hard in general non-equilibrium
systems.  However, non-equilibrium properties can be inferred by other
means: suitable asymmetric time correlation functions can be
constructed to unmask the breaking of time reversal symmetry providing a
proxy for the departure from equilibrium
\cite{Pomeau,Joss,FalkovichPnas}. Looking at the time evolution of
response functions, which describe how some system variables relax
to their statistically steady state after a perturbation on the same or
different degrees of freedom, can reveal aspects of the asymmetries
between degrees of freedom induced by the presence of currents
\cite{Zwanzig,FDreport,sarra2021}.

Fluid flows maintained by an external supply of kinetic energy, acting
at large scales, which is dissipated into heat at small scales by
viscous forces are characterized by a turbulent-state which is an important example of non-equilibrium statistically steady state
\cite{RoseSulem78,Frisch95,falkovich2006lessons}. In such a
turbulent-state, a current of energy flows from the large to the small
scales thanks to the nonlinearity of the Navier-Stokes equation (NSE),
with a constant  (on average) flux across the scales -- after
Richardson this is dubbed the direct energy cascade scenario
\cite{Frisch95}. Typically the energy cascade is studied in terms of
single time statistical objects, e.g. the third order moment of the
velocity differences between points at distance $r$ is directly linked
to the energy flux via the celebrated $4/5$ law \cite{Frisch95}, which is one
of the few exact results that can be obtained on the turbulent-state
and entails a spatial asymmetry in the statistics of the fluctuations
of the velocity field. Much fewer studies attempted a direct study of
the energy cascade, and the consequential asymmetries, with reference
to the time evolution of the flow \cite{Joss,FalkovichPnas}.

In this paper, we aim at studying the temporal properties of the
turbulent energy cascade in terms of two statistical tools: asymmetric
time correlations and response functions. We apply these tools to
shell models \cite{bohr1998,biferale2003,ditlevsen2010} that are
relatively (with respect to NSE) low dimensional dynamical systems that
phenomenologically (and to some extent quantitatively) reproduce the
main features of the turbulent energy cascade. Such models are
constructed without a spatial structure using a discrete number of
Fourier shells with an associated complex variable, representing the
velocity fluctuation at that scale (inverse of wavenumber). This simplified
structure makes, in comparison to real fluid flows, the study of the
temporal properties easier while maintaining the main phenomenological
features of the problem
\cite{ottinger1996relation,biferale1999multi,CBBDP}. In particular, we
consider the so-called Sabra shell model \cite{Sabra} and study
suitable asymmetric time correlations of the energy at a given
shell and response functions to energy perturbations. Specifically,
concerning the latter, we test \textit{non-diagonal} responses,
i.e. we study how a perturbation of the energy at a given shell alters
the relaxation of the energy of neighboring shells both in and against
the direction of the energy flux. Previous studies in the shell model
\cite{FDTturb, Japs2021} focused on diagonal response functions
considering perturbations of the shell velocity. While such a
procedure is interesting with respect to the fluctuation dissipation
relations \cite{kubo1966fluctuation,FDreport,lucarini2012beyond}, it is not able to reveal the asymmetries induced by
the cascade, which are, instead, clearly detected when perturbing the
shell energy and looking at non-diagonal response functions.

At first we compare the behavior of the asymmetric time correlations
and response functions either when the system is forced and dissipated
or unforced and inviscid.  In the latter case, unlike the former, an
equilibrium state establishes similarly to the well known absolute
equilibrium of the truncated Euler equations
\cite{kraichnan1973helical,RoseSulem78}, which were numerically
studied by looking either at the transient stages leading to the
equilibrium state \cite{cichowlas} or at the spatio-temporal
decorrelation of two copies of the system \cite{murugan2021prl}. The
inviscid shell model represents a toy model version of the truncated
Euler equations. The asymmetric time correlations of the energy vanish
for the inviscid (equilibrium) shell model while are clearly different
from zero in the forced and dissipated (non-equilibrium) shell model,
quantifying the breaking of time reversibility. In particular, we show
that it is enough to look at the short time behavior of such
correlations which is associated to the third order moment of the
shell energy rate of change, similarly to what observed in
Ref.~\cite{FalkovichPnas} for tracer dynamics.  Similarly, clear
asymmetries in the relaxation of energy at wavenumbers smaller or
larger than the perturbed shell distinguish the non-equilibrium and
equilibrium cases and the direction of the cascade in the former case.
Then, we apply these tools to the case in which the shell model is
forced at intermediate scales. At scales smaller than the forced one
the usual energy cascade is expected while, as we will show, at larger
scales the behavior of correlations and responses is compatible with
an equilibrium state as conjectured \cite{Frisch95} and to some extent
shown also in direct numerical simulations of NSE
\cite{dallas2015statistical,alexakis2019thermal} and experiments
\cite{gorce2022statistical}, although this was recently challenged
\cite{ding2023departure,hosking_schekochihin_2023}.

The paper is organized as follows. In Sec.~\ref{sec:model} we
introduce the shell model for turbulent energy cascade and make a
resume of its main properties.  Section~\ref{sec:tools} presents the
main tools used to probe the non-equilibrium properties of the energy
cascade, namely the asymmetric correlation functions and the
non-diagonal energy response functions. In Sec.~\ref{Results} we
present the results, in particular: first we contrast the behavior of
the asymmetric correlations of the shell energy in the shell model
forced at large scales with those of the inviscid and unforced shell
model; then we explore the non-diagonal energy response functions in
the same settings; finally we consider the forced shell model and
explore the behavior of the two quantities at scales larger and
smaller than the forcing scale in order to ascertain the equilibrium
or non-equilibrium character of the former.  Section~\ref{Conclusions}
is devoted to conclusions and offers a perspective on future
investigations. Several Appendices complement the main text: 
App.~\ref{app:numerics} provides some details on the simulations
including a table with the parameters; App.~\ref{app:Pomeau} discusses
some subtleties related to the computation of the asymmetric
correlation functions; App.~\ref{app:equilibrium} complements the study
of the inviscid unforced shell model; finally, App.~\ref{appendixdRdt}
provides a derivation for the small time behavior of the energy
response functions.

\section{Model\label{sec:model}}
Shell models are finite-dimensional dynamical systems designed to
reproduce the phenomenology of the turbulent energy cascade
\cite{bohr1998,biferale2003,ditlevsen2010}.  The basic idea is to
consider a discrete set of wavenumber $k_n\!=\!k_0 2^{n-1}$, the shells with index $n\!=\!1,\ldots,N$. A single complex
variable, $u_n$, is used to represent the velocity fluctuations at
scale $k_n$. The velocity variables $\{u_n\}_{n=1}^{N}$ evolve with a
set of equations formally analogous to the Navier-Stokes equation in
Fourier space:
\begin{eqnarray}
\dot{u}_n\!=ik_n Q[u,u] \!-\!\nu k_n^2u_n\!+\!f_n\,,   \label{eq:shell}
\end{eqnarray}  
where  $\nu$ is
the viscosity and $f_n$ the forcing. The quadratic term $Q(u,u)$ is
built to ensure that in the unforced and inviscid limit ($f_n\!=\!\nu\!=\!0$),
as for the three-dimensional Euler equation, the dynamics preserves both energy $E$ and
helicity $H$, that, for the shell model, read
  \begin{eqnarray}
    \label{eq:E}
      E &=& \sum_{n=1}^{N} e_n\,,\\
     \label{eq:H}
    H &=& \sum_{n=1}^{N} (-1)^{n} k_n e_n\,,
  \end{eqnarray} 
where $e_n=|u_n|^2/2$ is the energy of shell $n$.
Yet there is some freedom in choosing the quadratic term $Q(u,u)$;
here, we consider the Sabra model \cite{Sabra}, for which
\begin{eqnarray}
Q(u,u)=2u_{n+2}u_{n+1}^*\!-\!\tfrac{1}{2}u_{n+1}u_{n-1}^*\!+\!\tfrac{1}{4}u_{n-1}u_{n-2}\,, \label{eq:sabra}
\end{eqnarray}  
where $^*$ denotes
complex conjugation,
with boundary conditions
$u_{-1}\!=\!u_{0}\!=u_{N+1}\!=\!u_{N+2}\!=\!0$.  The choice to restrict the quadratic
interactions to neighboring shells is justified by the idea that the
energy cascade is \textit{local} in scale \cite{RoseSulem78}, i.e. the
energy transfer is mainly due to the interaction with close-by scales.  As demonstrated in
many studies (see, e.g.,
Refs.~\cite{bohr1998,biferale2003,ditlevsen2010}), shell models,
including the Sabra \cite{Sabra}, display the same phenomenology of the
turbulent energy cascade.

The forcing $f_n$, usually localized around some scale $1/k_{n_f}$
($f_n\neq 0$ for $n\sim n_f$), injects energy at a rate
$\epsilon=\sum_{n} \langle \Re\{f_nu_n^*\}\rangle$, where $\langle
\cdot\rangle$ denotes time averages and $\Re\{\cdot\}$ the real part. The nonlinear term transfers
the energy at smaller scales, on average, with a constant flux equal
to $\epsilon$. And, at large enough wavenumber, the viscous term
becomes important and removes the energy at a rate $\nu \sum_{n} k_n^2
\langle |u_n|^2\rangle=\nu \Omega=\epsilon$ ($\Omega$ being the
enstrophy). In this way, a non-equilibrium steady state is
established, which can be described by the energy balance equation
\begin{equation}
\dot{E}_M= - \Pi_{M}-\nu \Omega_M +  \sum_{m=1}^M \langle \Re\{f_mu_m^*\}\rangle\ ,
\label{eq:energybalance}
\end{equation}
where $E_M=\sum_{m=1}^{M} \langle e_m\rangle$ and
$\Omega_M=\sum_{m=1}^{M} k_m^2 \langle e_m\rangle$ ($M\leq N$) are the
average energy and enstrophy up to wavenumber $M$, respectively; while
\begin{equation}
  \Pi_M\!\!=\! \Delta_{M+1}\!+\!\tfrac{1}{2}\Delta_{M}\,,\quad
  \Delta_m\!\!=\!k_m\Im\{u_{m+1}u^*_{m}u^*_{m-1}\}\,,
  \label{eq:flux}
\end{equation}
$\Im\{\cdot\}$ denoting the imaginary part, is the energy flux transferred to shells $>M$ (\textit{viz.} minus the
rate of energy loss from shells $\leq M$) due to the nonlinear
terms. Notice that $E_N=E$ and $\Pi_N=0$, since the non-linear term
preserves energy, so that when summing over all the shells
$\dot{E}=\epsilon-\nu\Omega=0$, i.e. energy is conserved on average.
For shells $m$ between the forced one ($n_f$) and that where dissipation
becomes effective the average energy flux is constant $\Pi_m=\epsilon$
(as $\dot{E}_m=0$ for each $m$ at stationarity), which is the hallmark
of the energy cascade. From $\Pi_m=\epsilon$ and (\ref{eq:flux}) one
can dimensionally see that $u_n \sim (\epsilon k_n^{-1})^{1/3}$ which
is the Kolmogorov 1941 \cite{Frisch95} prediction for the shell model,
which allows to estimate the dissipative wavenumber as $k_{n_d}\approx
(\epsilon/\nu^{3})^{1/4}$.

Remarkably, shell models also reproduce some quantitative aspects of
the statistical properties of turbulent flows as, for instance, the
anomalous scaling of the velocity structure functions (moments of
velocity differences), which can be expressed as $\langle
|u_n|^p\rangle$ and, in the inertial range ($n_f<n<n_d$), are found to
scale as
\begin{equation}
  S_p(k_n)=\langle |u_n|^p\rangle \sim k_n^{-\zeta(p)}
\label{eq:sf}
\end{equation}  
with $\zeta(p)$ very close to the exponents measured in turbulent
flows, and thus deviating from the Kolmogorov dimensional prediction
$p/3$.

In the inviscid and unforced case ($\nu=f_n=0$), the system sets on an
equilibrium statistically stationary state while evolving on a
manifold dictated by the value of the energy and helicity, akin to the
one of truncated Euler equations \cite{kraichnan1973helical,
  RoseSulem78} (see also
  Refs.~\cite{cichowlas,murugan2021prl}).  Thus, the SABRA model does
offer a nice laboratory to contrast the equilibrium and
non-equilibrium properties of the Euler and Navier-Stokes equations,
which is the aim of this paper.

We will consider the Sabra model
forced at large scales and contrast it with the unforced inviscid case
and, also, the Sabra model forced at intermediate scales. In the
latter case one expect that at scales above the forcing the physics
will be akin to that of the inviscid case (i.e. equilibrium) and below
the forcing scale characterized by a direct energy cascade
(i.e. non-equilibrium)
\cite{dallas2015statistical,alexakis2019thermal,gorce2022statistical,alexakis2023fluctuation}.

We conclude this section mentioning that in the rest of the paper we
shall present simulations obtained by using a forcing which imposes a
fixed energy input $\epsilon=1$. However, we tested that the results remain
qualitatively unchanged by using a constant forcing on a single
shell. Appendix~\ref{app:numerics} provides other details on the
numerical implementation and the parameters which have been used.

\section{Probing the non-equilibrium properties of the turbulent cascade
  \label{sec:tools}}
Let us now present the main statistical mechanics tools we
used to characterize the non-equilibrium properties of the energy
cascade, namely: asymmetric correlation functions, which allow for
detecting irreversibility \cite{Pomeau,Joss,FalkovichPnas}, and
response functions \cite{FDreport}, which by describing how
observables at a given shell $n$ behave after a suitable perturbation
is performed at shell $m$ allow for highlighting the asymmetries between
shell variables induced by the energy flux.

\subsection{Testing the breaking of time-revesal symmetry via asymmetric correlation
  functions\label{sec:ToolsPomeau}}

The hallmark of out-of-equilibrium systems is the breaking of the
time-reversal symmetry that, mathematically, stems from the absence of
detailed balance with the associated positive entropy production
\cite{ruelle1996positivity,seifert2012stochastic,peliti2021stochastic}.
However, entropy production is a global quantity difficult to measure.
Here, to detect temporal asymmetries, we take an alternative route by
looking at the behavior of suitable correlation functions.  Indeed, as
made clear e.g. by Onsager \cite{onsager1931reciprocal}, for any choice of
observable functions $f$ and $g$ of the system state, at equilibrium
one has $\langle f(t)g(0)\rangle = \langle f(0)g(t)\rangle$ due to
time reversibility. This result entails that if there exist $f$ and
$g$ such that $\langle f(t)g(0)\rangle \neq \langle f(0)g(t)\rangle$
the system is out-of-equilibrium, and the difference $\langle
f(t)g(0)\rangle-\langle f(0)g(t)\rangle$ can be taken as a proxy of
the distance from equilibrium. The functions $f$ and $g$ could refer
to different observables of the system, e.g. different dynamical
variables, or if the same observable is used one must consider
appropriate functions thereof.

Here, following Refs.~\cite{Pomeau,Joss}, after denoting with $x(t)$ a
statistically stationary signal representing the temporal evolution
of an observable of the system, we will consider asymmetric
time-correlation functions of $x$ to detect breaking of time reversal
and thus the signature of non-equilibrium. In particular, we consider
the function  \cite{Joss}
\begin{equation}
    \Psi_x(\tau) \equiv \langle x^2(t)x(t+\tau) \rangle - \langle
    x(t)x^2(t+\tau) \rangle\,, \label{eq:psi}
\end{equation}
which, for a stationary signal, can be equivalently written as 
\begin{equation}
    \Psi_x(\tau) = \tfrac{1}{3}\langle [x(t+\tau) - x(t)]^3
    \rangle\equiv\Phi_x(\tau)\ .
    \label{eq:phi}
\end{equation}
This latter expression was used in \cite{FalkovichPnas} to reveal
the irreversibility of tracers dynamics in turbulence. While
mathematically $\Phi_x(\tau)=\Psi_x(\tau)$, in numerical computation of
Eqs.~(\ref{eq:psi}) and (\ref{eq:phi}) some differences may appear due to
unlike statistical convergence, especially at small time lags. Such
differences, as discussed in Appendix~\ref{app:Pomeau}, if not
properly considered may lead to spurious results.

In particular, in the case of the shell model, we shall compute the
above introduced correlation function using as observable the
instantaneous energy at shell $n$, i.e. $x(t)=e_n(t)=\frac{1}{2}
|u_n(t)|^2$, and consider different shells.  As clear from
Eq.~(\ref{eq:phi}), this choice is expected to provide also some hints
on the energy cascade process as it informs us about how the energy of
a shell varies in time. We notice that taking as $x(t)$ the signal of
the real or imaginary part of the shell velocity leads to results
essentially indistinguishable from equilibrium. This means that it is
important to use proper observables.

\subsection{Testing asymmetries among the degrees of freedom via non-diagonal  response functions\label{sec:ToolsRF}}
One of the most distinguishing aspects of out-of-equilibrium systems
is the presence of currents, e.g. of matter or energy
\cite{mazur84}. Besides breaking the time-reversal symmetry, as
discussed in the previous section, these currents generate asymmetries
in the degrees of freedom of the system. In the specific case of
turbulence, energy flows from the injection scale towards smaller
scales until it is dissipated at very small ones. Thus there is a
wide range of scales, which increases with the Reynolds number
\cite{falkovich2006lessons}, interested by such an asymmetry.  In
other systems the lack of symmetry may lead to spatial inhomogeneity
between the degrees of freedom \cite{sarra2021}.

A powerful statistical mechanics tool to explore such asymmetries is
the study of response functions (RFs) \cite{FDreport}. The idea is
rather simple, if we consider generic time-dependent observables
having $M$ components with $\bm A(t) = \{ A_m(t) \}_{m=1, \dots, M}$
and $\bm B(t) = \{ B_n(t) \}_{n=1, \dots, M}$, and we denote with $\bm
A'(t)$ and $\bm B'(t)$ their trajectory after a perturbation of one of
the components of $\bm A$, then the usual definition of (impulsive) response
function reads:
\begin{equation}
    R_{A_m, B_n}(t) = \frac{\overline{B'_n(t)-B_n(t)}}{A'_m(0)-A_m(0)}
    \equiv \frac{\overline{\delta B_n}(t)}{\delta A_m(0)}\ ,
\end{equation}
where $\delta [\cdot]$ is the difference between the value in the
perturbed system and that in the unperturbed one, while
$\overline{[\cdot]}$ denotes the statistical average over many different
realizations of the experiment. The above formula measures how much an instantaneous perturbation on $A_m$ influences, on average, the variable $B_n$ at
later times.

In the case of the shell model a natural observable to perturb is the
energy at a given shell $e_m(t)=\frac{1}{2} |u_m(t)|^2$ and then to
look at the effect of the perturbation on the energies at shells
larger (scales smaller) and smaller (scales larger) than $m$. Owing to
the energy flux a clear asymmetry between the two directions should be
expected, while it should be absent in the equilibrium case, i.e. in
the inviscid and unforced limit. To the best of our knowledge previous
attempts to measure the response functions in the shell model
\cite{FDTturb,matsumoto2014response} only considered diagonal
responses with respect to infinitesimal perturbations on the shell
velocity $u_m$ at a given scale. Here, the main novelty is to perturb
the energy by a non-infinitesimal amount (somehow similarly to
Ref.~\cite{Rfinpert}) and to explore non-diagonal responses, $R_{e_m,e_n}(t)$
($n\neq m$). In the following, we shall denote such
response function as $R_{m,n}(t)$ to ease the notation.

In principle, it is not needed to physically perturb a system in order
to study the response functions. In the linear perturbation
regime, RFs are linked, via the
fluctuation-dissipation relations (FDRs), to suitably defined
correlation functions
\cite{kubo1966fluctuation,FDreport,lucarini2012beyond}.  As an
important example, we mention the Green-Kubo formulas
\cite{kubo1966fluctuation} which link the response to an external
field with correlations computed at equilibrium.  FDRs, originally derived 
for equilibrium Hamiltonian systems, apply
also to non-equilibrium systems as any response can be expressed
as correlation functions whose functional form depends on the
invariant measure of the system
\cite{falcioni1990correlation,ruelle1998general}.  It is worth
stressing that there is still a certain confusion in the literature on
this aspect. Indeed,  FDRs hold under very general
assumptions and are only slightly more complicated for dissipative
chaotic systems. Such complications, which are independent of the
equilibrium or non equilibrium nature of the system, arise due to the
singular character of the invariant measure. However, with the
addition of a small amount of noise in the dynamical equations
(i.e. smoothing the measure) it is still possible to compute RFs via some
correlation function
\cite{harada2005equality,harada2006energy,baldovin2021handy}.  This
idea has been exploited, e.g., in the context of the shell model in
\cite{matsumoto2014response}.

In general, for all systems which cannot be directly probed via
perturbations (the climate is in this respect an example of utmost importance
\cite{lucarini2017predicting,ghil2020physics}) it is crucial to
exploit FDRs to understand the possible effect of local disturbances.
However, in our study we shall directly study the energy RFs, since from a computational point of view exploiting
FDRs as in
Refs.~\cite{harada2005equality,harada2006energy,baldovin2021handy}
would anyway require a modification of the equation with the addition
of noise and a huge statistics in order to allow for
cancellations in the correlation functions
\cite{matsumoto2014response}.
\begin{figure*}[t]
\centering
\includegraphics[width=0.9\textwidth]{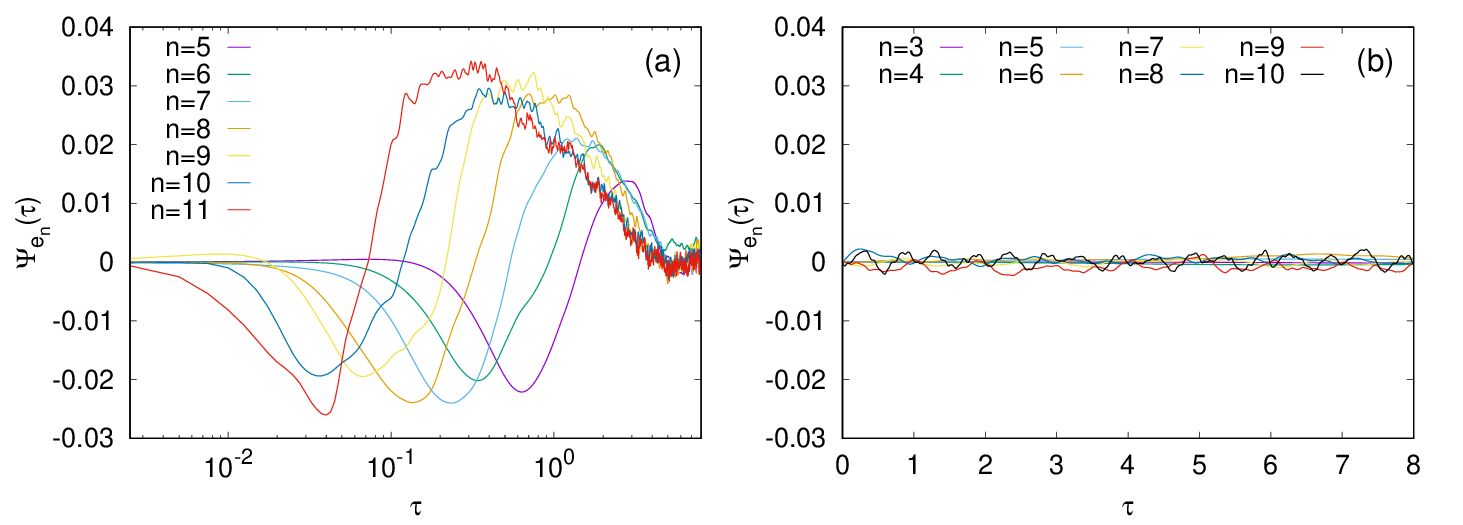}
\caption{Asymmetric time-correlation functions $\Psi_{e_n}(t)$ at
  varying $n$ as labeled. (a) Turbulent case, shells $n$ in the
  inertial range. Different functions show a qualitatively similar
  behavior, and symmetry under time reversal is absent. Statistics:
  $6\cdot 10^{5}$ samples. (b) Inviscid case, shells $n$ as labeled in
  the range with energy equipartition. Small fluctuations around zero
  are compatible with $\Psi_{e_{n}}(t)=0$. Statistics: $5\cdot 10^{5}$
  samples.}
\label{fig:Pomeau_Turb_Eq}
\end{figure*}

\section{Results\label{Results}}
We present now the results obtained with numerical
simulations performed on the Sabra shell model, of both inviscid
and viscous type. We will highlight the differences between a system
at equilibrium (the former) and a system out-of-equilibrium (the
latter). We shall first discuss the test of irreversibility
(Sec.~\ref{sec:ToolsPomeau}) and the response-functions
(Sec.~\ref{sec:ToolsRF}) for the shell model either forced at large
scales (run-LSF) or unforced and inviscid (run-Eq), for the parameters
see Table~\ref{tab:param} of App.~\ref{app:numerics}. Then we shall
discuss the shell model forced at intermediate scales (run-ISF),
showing that the physics at scales larger than the forcing scale is
akin to the inviscid model (equilibrium) while at scales smaller than
the forcing is that of the non-equilibrium energy cascade.

\subsection{Irreversibility\label{sec:ResIrreversibility}}
As discussed in Sec.~\ref{sec:ToolsPomeau} in order to test the
irreversibility of the energy cascade we focus on the energy signal
$e_n(t)=|u_n|^2/2$ at shell $n$ (by varying $n$). In
particular, we measure the asymmetric correlation function $\Psi$
(\ref{eq:psi}) or, equivalently, $\Phi$ (\ref{eq:phi}), which we
rewrite as
\begin{eqnarray}
&& \Psi_{e_n}(\tau)=\frac{\langle e_n^2(t)e_n(t+\tau) \rangle -
    \langle e_n(t)e_n^2(t+\tau) \rangle}{\langle e_n^3(t)
    \rangle}=\nonumber\\  = && \Phi_{e_n}(\tau) =
  \frac{1}{3}\frac{\langle [e_n(t+\tau)-e_n(t)]^3\rangle}{\langle
    e_n^3(t) \rangle}\,, \label{psi_En}
\end{eqnarray}
to simplify the notation we have used the same symbol, even if the two
quantities are now made nondimensional. We also recall that the two
different expressions should coincide for a stationary signal, but can
be numerically different for finite statistics (see
App.~\ref{app:Pomeau}).

The inviscid and unforced shell model reaches an equilibrium
statistically stationary state with no net currents and, thus,
detailed balance holds, implying that forward and backward dynamics
are indistinguishable. As a consequence we should expect that
$\Psi_{e_n}(\tau)=\Phi_{e_n}(\tau)=0$. Conversely, for the forced and
dissipated shell model, owing to the presence of an energy flux from
the large to the small scales, the dynamics is irreversible and
$\Psi_{e_n}(\tau)\neq 0$.  Such expectations are well verified as
shown in Fig.~\ref{fig:Pomeau_Turb_Eq}: apart from unavoidable
fluctuations, $\Psi_{e_n}(\tau)=0$ for the inviscid model
(Fig.~\ref{fig:Pomeau_Turb_Eq}b), while for the forced model is
clearly different from zero (Fig.~\ref{fig:Pomeau_Turb_Eq}a) and,
actually, it displays several interesting features.
\begin{figure}[t]
\centering
\includegraphics[width=0.48\textwidth]{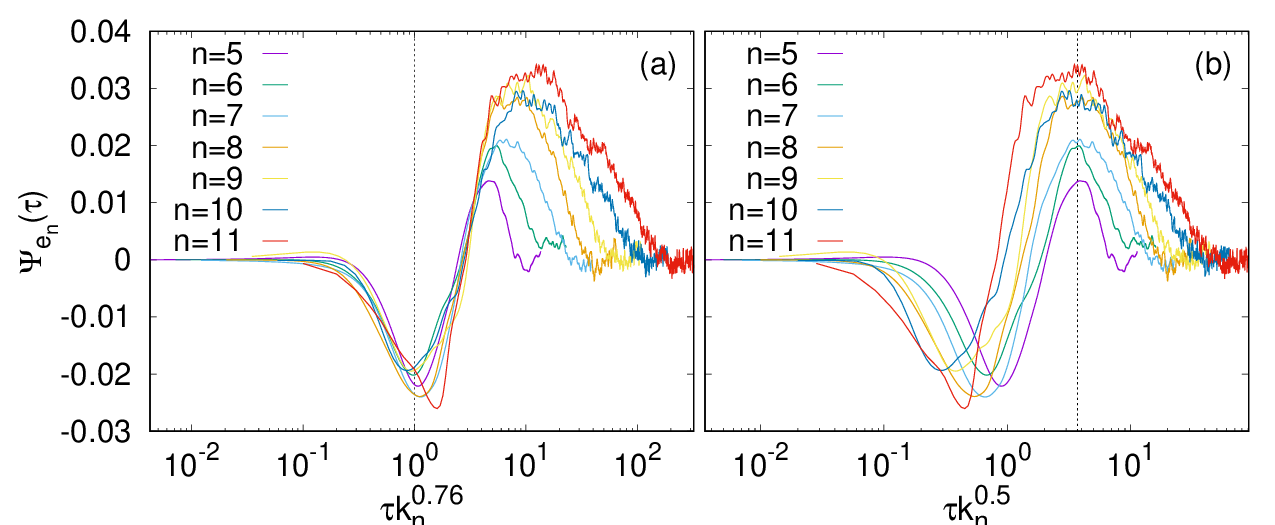}
\caption{Same as Fig.~\ref{fig:Pomeau_Turb_Eq}(a) with time rescaled by (a)
  $\tau_n=k_n^{-0.76}$ (see text for a discussion) and (b) $k_n^{-0.5}$,
  where $0.5$ was chosen in such a way to line up the maxima. The
  vertical dashed lines are drawn to guide the eyes making clearer the
  line up. }
\label{fig:Pomeau-rescaled}
\end{figure}
\begin{figure*}[t!]
  \centering
  \includegraphics[width=0.9\textwidth]{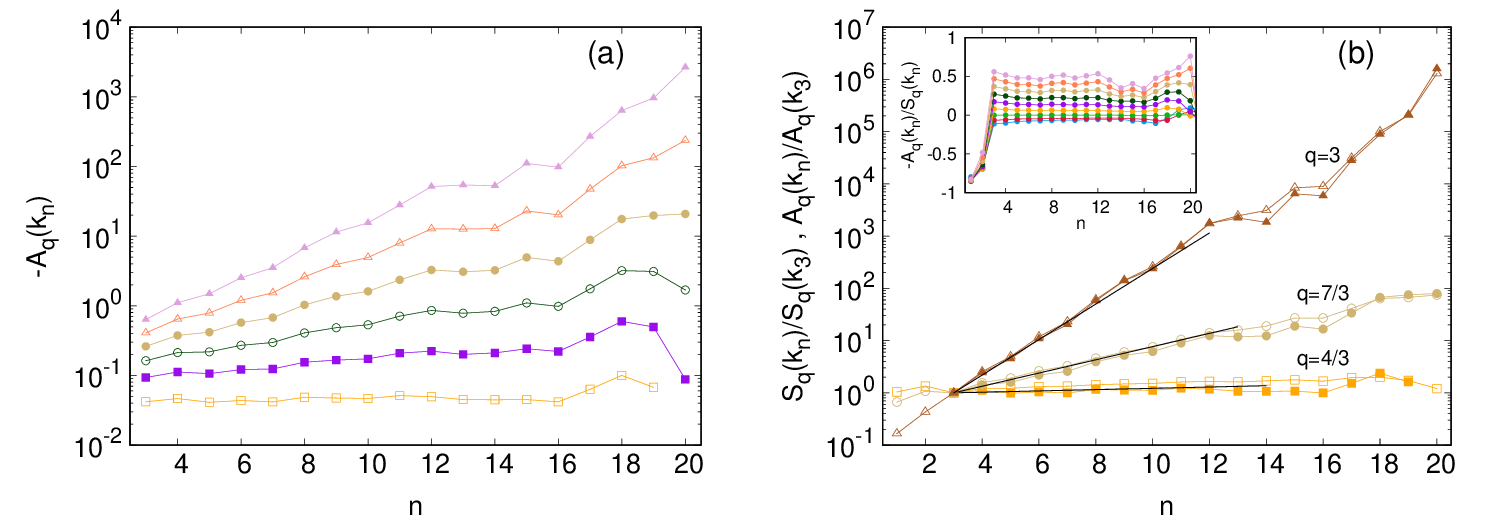}
  \caption{ (a) Asymmetric moments $-\mathcal{A}_q(k_n)$
    (Eq.~\eqref{eq:asymm}) vs the shell number $n=\log_2(k_n/k_0)$,
    for increasing $q=s/3$ with $s=4,\ldots,9$ (from bottom to top).
    The moments for $q<1$ are positive (not shown) and
    $\mathcal{A}_1(k_n)=0$ by stationarity.  The lowest and highest
    shells are omitted. (b) Symmetric moments $S_q(k_n)$
    (Eq.~\eqref{eq:simm}) (empty symbols) and asymmetric ones
    $-A_q(k_n)$ (filled symbols) rescaled by their value at $n=3$ for
    three values of $q$, as labeled.  The black solid lines display
    the prediction \eqref{pnMF}. Inset: the ratio
    $-A_q(k_n)/S_q(k_n)$, with $q=s/3$ for $s=1,\ldots,9$ from bottom
    to top.  Statistics is over $2.5 \cdot 10^6$ samples.}
    \label{fig:A_q}
\end{figure*}


First of all, for small time lags $\Psi_{e_n}(\tau)$ is negative while
it becomes positive at larger time lags and then it approaches zero at
much larger $\tau$'s. This is consistent with the idea that energy
decreases on a fast time scale and increases over a longer time scale
(notice that by stationarity $\langle e_n(t+\tau)\rangle =\langle
e(t)\rangle$). This is similar to the flight-crash events described in
Ref.~\cite{FalkovichPnas} but at the level of a single shell. The fact
that the energy loss is faster than the energy gain is physically
understood from the fact that the time scale of the shell decreases
with the scale as $\tau_n\sim (u_nk_n)^{-1}\sim k_n^{-2/3}$ (where we
used Kolmogorov scaling). Therefore, it takes longer to receive energy
from larger and slower scales (smaller $n$) than to transfer it to
smaller scales (larger $n$). However, as already clear from
Fig.~\ref{fig:Pomeau_Turb_Eq}a, in the temporal behavior of
$\Psi_{e_n}(\tau)$ several time-scales are at play.  Indeed one can see
that the short time minima depends on the shell index $n$ while the
large-time decay of the curves is essentially independent of $n$, as
the curves reach zero at about the same time. The presence of many
time scales is even clearer by looking at
Fig.~\ref{fig:Pomeau-rescaled}, showing that different rescaling of
the time lags depending on the scale as $k_n^{\beta}$ are needed to
make either the minima ($\beta\approx 0.76$,
Fig.~\ref{fig:Pomeau-rescaled}a) or the maxima ($\beta\approx 0.5$,
Fig.~\ref{fig:Pomeau-rescaled}b) line up. While we have no clear
understanding of the latter exponent, the first one can be understood
as follows.

As discussed in the next section and Appendix~\ref{app:Pomeau}, at
short times it should hold
\begin{equation}
\Psi_{e_n}(\tau)=\Phi_{e_n}(\tau)\approx \frac{\langle \dot{e_n}^3\rangle}{\langle e_n^3\rangle} \tau^3\,.
\label{eq:scalingPsi}
\end{equation}
Equations~\eqref{eq:sf} and \eqref{pnMF} (see next section) imply that
${\langle \dot{e_n}^3\rangle}/{\langle e_n^3\rangle}\sim
k_n^{3-\zeta(9)+\zeta(6)}$.  From our simulations we get
$\zeta(6)\approx 1.70(2)$ and $\zeta(9)\approx 2.43(6)$.  If we now
re-scale $\tau$ in Eq.~\eqref{eq:scalingPsi} with $\tau_n\sim
k_n^{-\beta}$ by requiring that the expression does not depend on $n$
we have $\beta=(3-\zeta(9)+\zeta(6))/3\approx 0.76(3)$, which is well
compatible with Fig.~\ref{fig:Pomeau-rescaled}a.  We also notice that
$\Psi_{e_n}$ is the difference of two disconnected, single scale
and multi-time correlation functions that, as thoroughly discussed in
Ref.~\cite{biferale1999multi}, should involve a whole hierarchy of
fluctuating eddy-turnover times from the shortest up to the largest.
In Ref.~\cite{biferale1999multi} it was discussed the general
framework of multi-time and multiscale correlation functions,
similarly to $\Psi_{e_n}$: one can define suitable correlations
involving energy at different shells and different times, which besides
providing information on the irreversibility can also further
characterize the physics of the energy cascade. However, this goes
beyond the aim of this work and we leave it for possible future
studies.

\begin{figure*}[t]
  \includegraphics[width=\textwidth]{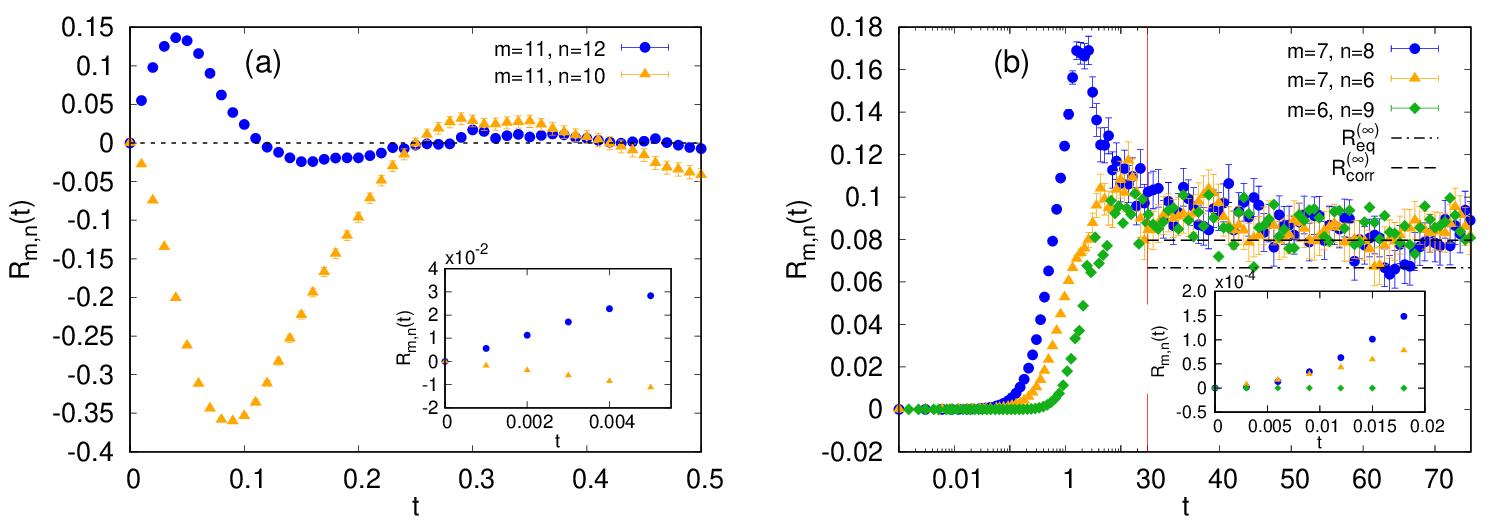}  
  \caption{Non-diagonal energy RFs $R_{m,n}$, with error bars,
    measured at the neighboring shells of the perturbed one. (a)
    Turbulent Sabra model. Initial energy perturbation: $\delta_{m=11}
    \simeq 9.93\cdot10^{-3}$.  (b) Inviscid Sabra model. The initial
    transient is plotted with logarithmic time axis up to the red
    vertical line, the later stages with a linear axis. Another RF has
    been added (green plot), with more distance between $m$ and $n$,
    in order to show the common asymptotic value. The dot-dashed line
    shows $R^{(\infty)}_{eq}$ which is the value that would be reached
    if perfect energy equipartition occurs, the dashed line shows
    $R^{(\infty)}_{corr}$ which is the asymptotic value taking into
    account the boundary corrections (see App.~\ref{app:equilibrium}
    and main text). Initial perturbation $\delta_{m=6} = \delta_{m=7}
    \simeq 1.79\cdot 10^{-3}$. The insets of both figures are enlargements
    of the initial-time range, showing respectively a non-zero and a
    zero first derivative at $t=0$. In all figures statistics is over
    $5\cdot10^5$ realizations.\label{fig:Resp_Turb-Eq_E}}
\end{figure*}

\subsubsection{Power fluctuations\label{shell power}}
The negativity of $\Psi_{e_n}(\tau)$ for small $\tau$ can be further
scrutinized by studying directly the $\tau\to 0$ limit.  Indeed by
expanding Eq.~\eqref{psi_En} for small $\tau$ (see
Appendix~\ref{app:Pomeau}, also for a discussion on the subtle
numerical aspects related to the short time behavior of $\Psi$ and
$\Phi$) it is easy to realize that the initial negativity means that
$\langle \dot{e}_n^3\rangle<0$, while we know by stationarity that
$\langle \dot{e}_n\rangle=0$: this means that the statistics of
$\dot{e}_n$ is negatively skewed, which confirms the fact that it is more
probable to lose than to gain energy on the short time.  This
observation was originally made in Ref.~\cite{FalkovichPnas} in the
context of irreversibility of Lagrangian trajectories in turbulence,
and further analyzed in \cite{CBBDP} in both direct numerical
simulations and in shell model version of Lagrangian motion. In
particular, in Ref.~\cite{CBBDP} it was introduced a set of symmetric
and asymmetric moments to probe the asymmetry of the
distribution. Here, following \cite{CBBDP}, we consider the
moments:
\begin{eqnarray}
  \mathcal{S}_q(k_n) &=& \langle |\dot{e}_n|^q \rangle / \epsilon^q \label{eq:simm}\\
  \mathcal{A}_q(k_n) &=& \langle \dot{e}_n|\dot{e_n}|^{q-1} \rangle / \epsilon^q
\label{eq:asymm}
\end{eqnarray}
where the normalization by $\epsilon^q$ is only to make the quantities
dimensionless.  Numerical simulations show that $\mathcal{A}_q(k_n)>0$
for $q<1$ and $<0$ for $q>1$, for $q=1$ is zero by stationarity.
Moreover, as shown in Fig.~\ref{fig:A_q}a, the asymmetric moments
display a power law behavior, $-A_q(n) \sim k_n^{\alpha(q)}$.  To
rationalize the exponents $\alpha(q)$ we can use dimensional analysis
in the spirit of the multifractal model \cite{Frisch95}. Noticing that
$\dot{e}_n$ is energy divided by time, we can assume that for each
shell one has to use its own characteristic eddy turnover time, that
dimensionally can be estimated as $\tau_n\sim 1/(|u_n|k_n)$, so that
\begin{equation}
 \langle \dot{e}_n^q\rangle \sim \langle |u_n|^{2q} k_n^{q}|u_n|^{q}\rangle \sim k_n^q S_{3q}(k_n)\sim k_n^{q-\zeta(3q)}\,,
    \label{pnMF}
\end{equation}
where we used \eqref{eq:sf} in the last two steps.  It should be noted
that this argument is purely dimensional, thus it applies both to the
symmetric and asymmetric moments.  It is worth stressing that it is
not obvious a priori that $S_q(k_n)$ and $-A_q(k_n)$ should scale in
the same way nor that the asymmetric moment can be guessed with a
dimensional argument, indeed it depends on cancellations which cannot
be controlled. However, Fig.~\ref{fig:A_q}b shows that $S_q(n)$ and
$-A_q(n)$ possess the same scaling behavior (see also the inset)
and agree with the prediction \eqref{pnMF}.

\subsection{Energy response functions\label{Response_functions}}

In order to detect the scale asymmetry between shells, caused by the
average energy flux from large to small scales, we study non-diagonal
RFs on the shell energies $e_n(t)$ (see also Sec.~\ref{sec:ToolsRF}):
\begin{equation}
    R_{m,n}(t) = \frac{\overline{\delta e_n}(t)}{\delta e_m(0)}\ ,
    \label{R_emen}
\end{equation}
the first index referring to the perturbed shell and the second to the
shell one looks at.  To fix the initial energy perturbation $\delta
e_m(0)$ of each experiment we introduce the following perturbation on the velocity at $t=0$:
\begin{equation}
    u_m = \sqrt{2e_m} e^{i\theta_m} \quad \longrightarrow \quad u_m' = \sqrt{2(e_m+\delta_m)} e^{i\theta_m}
    \label{uprime}
\end{equation}
where $\theta_m$ is the phase of $u_m$, and $\sqrt{2e_m}$ its modulus.
The constant $\delta_m$ quantifies  the magnitude of the initial energy perturbation while keeping the phase of $u_m$ fixed, and is such that: $\delta e_m(0) = \delta_m$. The phases are crucial for the energy transfer \cite{biferale2017optimal} that is why we keep them constant. 
Moreover, we consider a non-infinitesimal perturbation: specifically, we choose $\delta_m$
to be a finite fraction of the typical fluctuation of the shell energy, namely its standard deviation:
\begin{equation}
    \delta_m = f\sigma_{e_n} = \frac{f}{2} \sqrt{\langle |u_n|^4 \rangle - \langle |u_n|^2 \rangle^2}\ ,\label{deltam}
\end{equation}
in which $f\simeq0.2$ was used.

\begin{figure*}[t]
  \includegraphics[width=\textwidth]{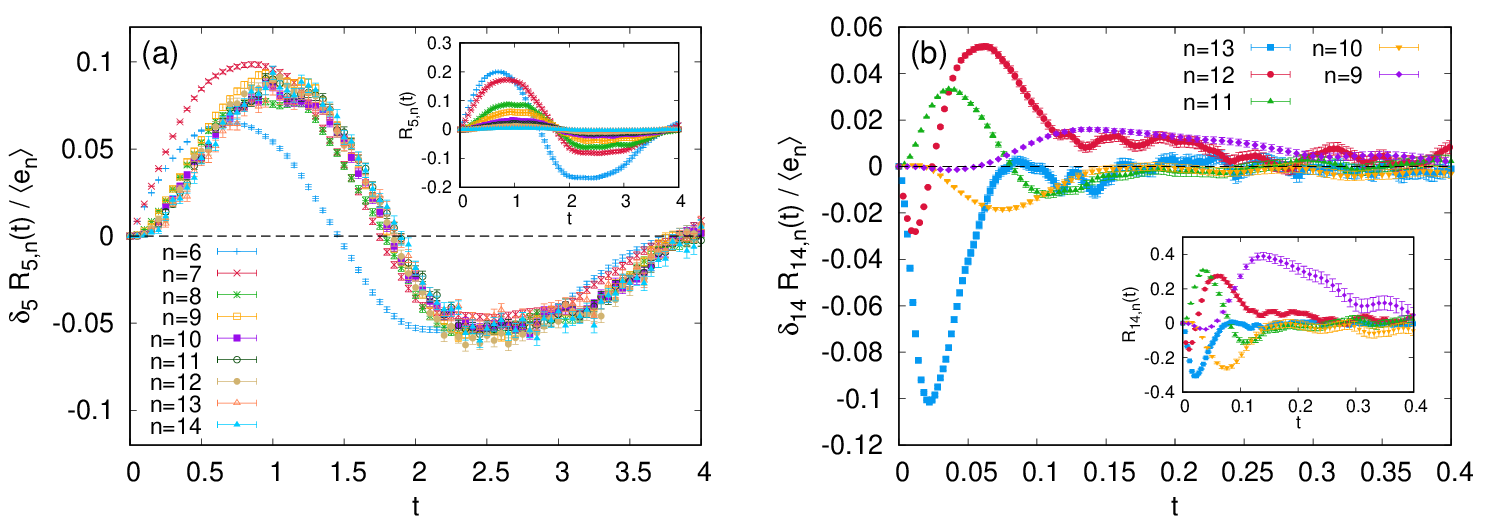}
  \caption{Relative energy deviation \eqref{eq:relative}
    at consecutive shells, with fixed perturbed shell $m$, in the turbulent shell model: (a) for shells $n>m=5$, with $\delta_{m=5} \simeq 0.1$; (b) for shells $n<m=14$ and $\delta_{m=14} \simeq 2.74\cdot10^{-3}$. The insets show the un-rescaled RFs.  Statistics is over $5\cdot10^5$ iterations. \label{Resp_Turb_E_forwbackw}}
\end{figure*}

In Fig.~\ref{fig:Resp_Turb-Eq_E} we compare the RFs $R_{m,n}$ with
$n-m = \pm 1$ in the turbulent (panel a) and inviscid (panel b).  The
functions in Fig.~\ref{fig:Resp_Turb-Eq_E}(a) clearly display the
asymmetry between degrees of freedom mentioned before: in the
turbulent system the opposite sign of the ``forward" (towards smaller
scales) and ``backward" (towards larger scales) RFs reveals the
presence of an overall energy current, displacing energy from larger
to smaller scales. The different amplitudes and relaxation times
relates to the fact that larger scales have larger amplitudes and are
slower than the smaller scales. On the other hand, such an asymmetry
is clearly lost in the inviscid system, as shown in
Fig.~\ref{fig:Resp_Turb-Eq_E}(b): the RFs are all positive, and after
an initial transient they approach a common non-zero asymptotic
value. This trend is explained as follows. In the inviscid system
energy (but also helicity) is conserved, therefore the energy
perturbation brings the system to a larger constant-energy
hyper-surface in phase space. Assuming a perfect energy equipartition
one can compute the expected long-time value
$R_{\text{eq}}^{(\infty)}=1/N$, however as discussed in
Appendix~\ref{app:equilibrium} boundary effects prevent perfect
equipartition. Taking them into account one can compute a corrected
asymptotic value, $R_{\text{corr}}^{(\infty)}$, which fits better with
the data as shown in Fig.~\ref{fig:Resp_Turb-Eq_E}(b)

It is equally interesting to notice that short shell-distance RFs show
non-zero initial derivatives in the turbulent case, whereas they are
zero in the inviscid system (see the insets of
Fig.~\ref{fig:Resp_Turb-Eq_E}). Briefly, this is due to the presence
or absence, respectively, of energy cascades in the system, since
these derivatives are related to the order-three correlator describing
the energy flux through the shells, as detailed in
Appendix~\ref{appendixdRdt}.

\begin{figure}[b!]
  \centering
  \includegraphics[width=.5\textwidth]{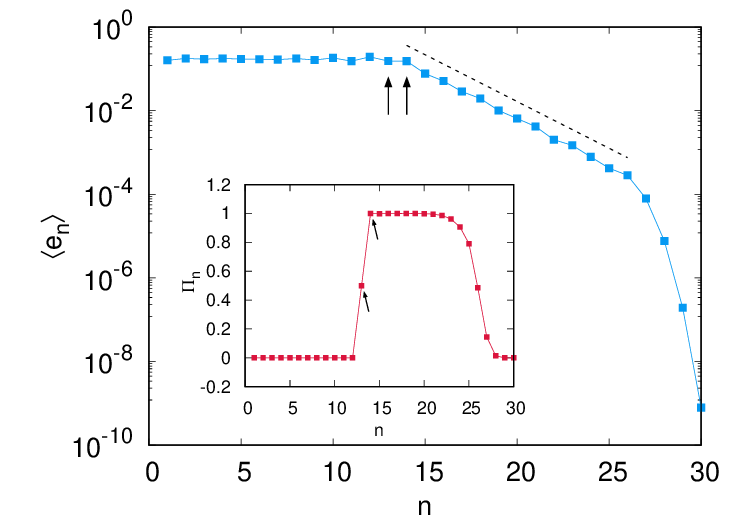}
  \caption{Energy spectrum $\langle e_n \rangle$ for the Sabra shell model with forcing at intermediate scales. The three  regimes of energy equipartition, power-law scaling and viscous damping can be identified. The dashed line corresponds to the scaling $\langle e_n\rangle \sim k_n^{-\zeta(2)}$, with $\zeta(2)\approx 0.74(4)$. Inset: average flux $\Pi_n$ \eqref{eq:flux} out of shell $n$. There is a clear transition from zero to positive outward flux. The small arrows in both figures indicate the forced shells. Statistics: $10^6$ samples.}
    \label{midforc_SF_FL}
\end{figure}
\begin{figure}[b!]
  \centering
  \includegraphics[width=.5\textwidth]{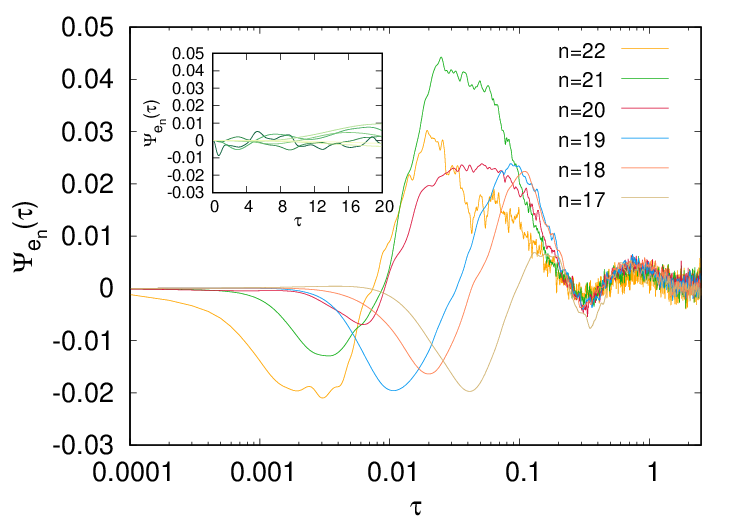}
  \caption{Asymmetric time correlation functions \eqref{psi_En} measured at shells larger than the forced ones, as  in label. Inset: the same correlations, measured at shells below forcing, $n$ ranging from 5 (lighter) to 11 (darker).  Statistics: $5\cdot 10^5$ samples.\label{midforc_Pom}}
\end{figure}

To offer a more complete investigation of the turbulent cascade we
also studied the response of shells further-apart from the perturbed
one.  Owing to the fact that the energy spectrum decays as a power
law, in the main panel of Fig~\ref{Resp_Turb_E_forwbackw}(a) we show
the relative energy deviation,
\begin{equation}
\frac{\delta_m}{\langle e_n\rangle} R_{m,n}= \frac{\overline{\delta e_n}(t)}{\langle e_n\rangle}\ , \label{eq:relative}
\end{equation}
for $n>m$. The advantage of the relative deviation (\ref{eq:relative}) is that
it allows for normalizing the amplitude of the response making possible the comparison of the response of shells at different distances from the perturbed one.
 Figure~\ref{Resp_Turb_E_forwbackw}(b) shows the
same for shells smaller than the perturbed one.  The insets represent
the usual (non-normalized) RFs. The energy deviations in the forward
direction collapse nicely onto the same curve, apart from the
functions with $n-m=1,2$. These two are the only functions whose index
$n$ is in the same wave-number triad of shell $m$, so a difference
with the other functions can be expected. On the other hand, the
energy deviations in the backward direction lack any kind of
similarity between themselves, and their short-time value, before the
relaxation, can be indeed positive (energy gain) or negative (energy
loss) for different values of $n-m$. Unlike the forward functions, the
backward ones show smaller amplitudes as the distance from the
perturbed shell grows. Overall, a correct interpretation of the latter
RFs requires a better understanding of how the different time-scales
involved, from the slowest to the one with shell index
$\text{max}\{m,n\}$, contribute to the energy deviation, in the same
way as assumed for multi-time multi-scale correlation functions
\cite{biferale1999multi}. And this is out of the scope of the present
work.
\begin{figure*}[t!]
	\centering
    \includegraphics[width=\textwidth]{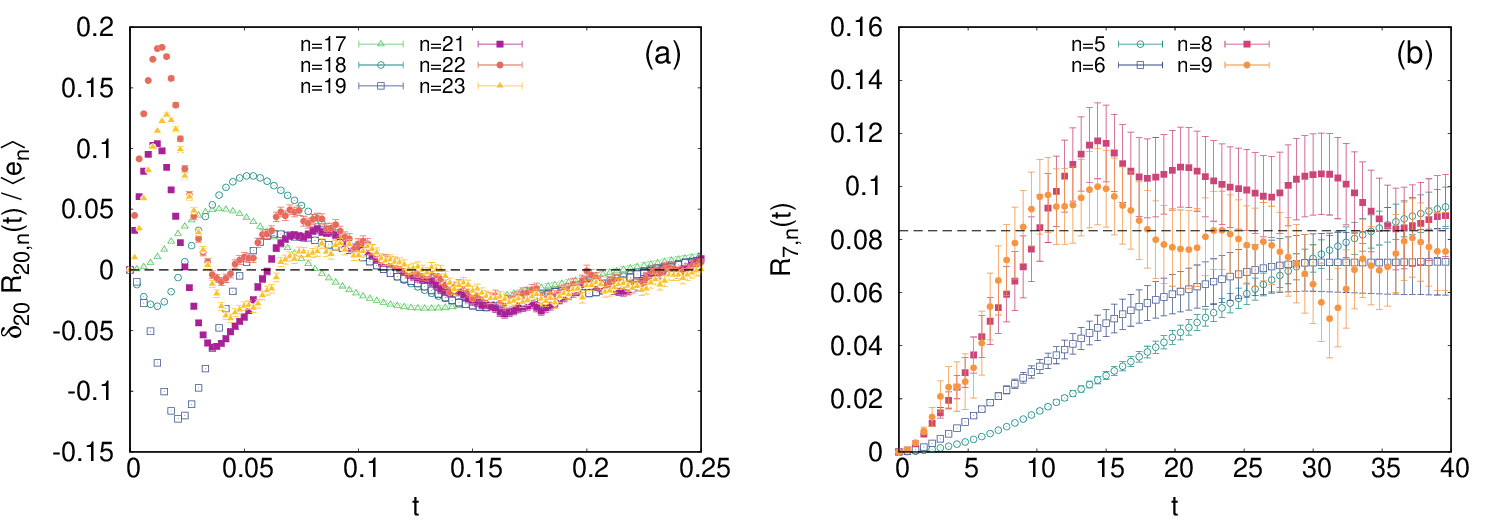}
    \caption{Energy response functions in the Sabra model with forcing
      at intermediate scales: (a) RFs normalized as in
      (\ref{eq:relative}) when perturbing shell $m=20$ in the inertial
      range. Both the forward ($n>m$) and the backward ($n<m$)
      functions are qualitatively similar to those of
      Fig.~\ref{Resp_Turb_E_forwbackw}a and b, respectively. (b)
      $R_{m,n}$ for $m=7$ in the middle of the equiparted part of
      the spectrum, above the forced shell. The RFs approach a common
      asymptotic value as in Fig.~\ref{fig:Resp_Turb-Eq_E}(b). The
      dashed line is estimated as $R_{eq}^{(\infty)}=1/N^*$, where
      $N^*$ is the number of shells for which energy equipartition
      holds. In both cases statistics is over $2\cdot10^5$
      realizations.    \label{midforc_resp}}
\end{figure*}

\subsection{Equilibrium/non-equilibrium at scales larger/smaller than the forcing scale}

So far we have discussed either the case of the inviscid, unforced
shell model or the forced shell model, showing how asymmetric energy
correlations or the energy RFs can reveal the asymmetries and breaking
of time reversal induced by the energy flux from the scale of forcing
to the smaller scales. It is now natural to wonder what does happen at
scales larger than the forcing scale. Since such scales are not
directly influenced neither by the forcing nor by the viscous damping
and owing to the direct cascade, it has been conjectured that the
physics of these large scales should be akin to the equilibrium one
\cite{Frisch95}. Numerical
\cite{dallas2015statistical,cameron2017effect,
  alexakis2023fluctuation, alexakis2019thermal} and experimental
\cite{gorce2022statistical} studies indeed seem to confirm that many
aspects of the scales above the injection scale are well captured by
absolute equilibrium theory \cite{kraichnan1973helical}. Two recent
studies \cite{ding2023departure,hosking_schekochihin_2023}, however,
seem to point to the fact that deviations from equilibrium can be
detected in direct numerical simulations. In particular, in
Ref.~\cite{ding2023departure} it is shown that the third order
velocity structure function is not zero (as a Gaussian-equilibrium
statistics would have prescribed) but decays as the second power of
the scale. This observation is substantiated by an inspection of the
Kàrmàn-Howarth-Monin equation \cite{Frisch95}, which ultimately shows
that such deviations can be ascribed to non-local interactions that,
however, are absent by construction in the shell model.  Furthermore,
the $k$-dependence of the energy spectrum at large scales appears to
be determined by the momentum injected by the forcing: for instance a
solenoidal, localized-in-space forcing would yield a large-scale
spectrum not compatible with absolute equilibrium
\cite{hosking_schekochihin_2023}.

From the above discussed works one can realize that it is interesting to investigate the
behaviors of asymmetric correlations and energy response functions at
scales smaller and larger of the forcing scale. To this aim we now
study the Sabra model with $N=30$ shells forced at intermediate scales
(i.e. $n_f=13$) so to have enough range of scales at shells larger and
smaller than the forced ones (see Appendix~\ref{app:numerics} for
details).

Figure~\ref{midforc_SF_FL} shows the energy spectrum, $\langle e_n
\rangle$, and the energy flux, $\Pi_n$, (see inset): energy
equipartition and zero energy flux obtained for shells smaller than
the forced ones are good indicators of statistical equilibrium. At
first we study the asymmetric correlation (\ref{psi_En}) to test
the time asymmetry of the energy evolution at scales below/above the
forcing one. As shown in Fig.~\ref{midforc_Pom}, $\Psi_{e_n}(\tau)$
for $n>n_f$ (main panel) is akin to the results of
Fig.~\ref{fig:Pomeau_Turb_Eq}(a) obtained with the usual large scale
forcing, while for $n<n_f$ (see inset) is statistically compatible
with zero, as expected in an equilibrium regime
(Fig.~\ref{fig:Pomeau_Turb_Eq}(b)).  Then we investigate the
relaxation of an energy perturbation, either perturbing a shell
$m>n_f$ (i.e. in the energy cascade range) or a shell $m<n_f$ (i.e. at
scales larger than the forcing scale). Figure~\ref{midforc_resp}(a)
displays the normalized RFs, i.e.  the relative energy deviations,
when the perturbed shell is larger than the forced ones, and their
behavior retain the features shown by both forward and backward
normalized RFs. Conversely, Fig.~\ref{midforc_resp}(b) refers to the
case where the perturbation acts on a shell smaller than forced ones:
as in the inviscid model, all RFs are positive and reach a common
asymptote whose value is again found, in first approximation, by
assuming energy equipartition among the degrees of freedom.

Summarizing, the behavior of both the asymmetric correlation and
the energy response functions gives strong evidence that, in the shell model, the
physics of the scales larger than the forcing one is compatible with
the equilibrium as in the inviscid shell model. We remark, however,
that there could be some properties which can deviate from
equilibrium. The dynamics clearly does not preserve neither the energy
nor the helicity and it is unclear what will be the effect on, e.g.,
the spectrum at the scales which match the forcing scales. As
discussed in Appendix~\ref{app:equilibrium}, in the inviscid case
oscillations in the spectra are expected at the boundary given by the
largest shell, this will be clearly modified by the presence of the
forcing. A confirmation of this difference is found in the good estimate of the asymptotic value computed with the perfect-equipartition assumption: boundary correction appears not necessary (Fig.~\ref{midforc_resp}(b)). In Ref.~\cite{ding2023departure} it was pointed out that
there are detectable deviations from Gaussianity at scales larger than
the forcing scale for the Navier-Stokes equations. We studied a
similar quantity, namely the third order velocity correlation function
(not shown), and we found it to vanish as also confirmed by the
analogous of the Kàrmàn-Howarth-Monin equation for the shell model. This
difference is likely due to the fact that the quadratic interaction
term is local in shell models.

\section{Conclusions\label{Conclusions}}
In this work we used asymmetric time correlation
functions and response functions to finite perturbations to ascertain
the breaking of time reversal symmetry and asymmetries between the
degrees of freedom in a simplified model for the energy cascade in
turbulence, i.e. the Sabra shell model
\cite{Sabra}.  We focused on the energy of a given shell and showed
that by looking at a suitable time correlation of the energy one can
clearly distinguish the case of a forced shell model displaying the
direct energy cascade of energy to the case of an unforced and
inviscid model. In the latter case the asymmetric correlation vanishes (meaning time reversible dynamics) while in the
former it is definitely different from zero and the behavior
understandable in terms of the Richardson energy cascade
scenario. Similarly, clear differences between equilibrium (inviscid-unforced case) and non-equilibrium (forced viscous case)
physics are clearly detectable using the energy response function, that
is, perturbing the energy at given shell and looking at the time
relaxation of the energy at a distant shell (either above or below the
perturbed one). A net difference is observed in the forced case
while looking at shells larger (scales smaller) or smaller (scales
larger) than the perturbed one, as a consequence of the average energy
current from large to small scales. Here the main novelty with respect
to previous studies has been to perturb directly the energy and not
the velocity.  The quantities studied in this paper thus allow to
observe (and, even if unnecessary, confirm) the celebrated Richardson
cascade scenario from the perspective of non-equilibrium Statistical
Mechanics. Finally, we considered the case of a shell model forced at
intermediate scales so to allow for scrutinizing the behavior of both
quantities at scales larger or smaller than the forcing scale. The
emerging figure seems to confirm the view that scales larger than the
forcing display properties which can be ascribed to the equilibrium
physics of the unforced-inviscid model.

There are at least two interesting perspectives emerging from our
study.  Within the context of simplified models it would be
interesting to explore, using the two tools we introduced, the shell
model with the nonlinear term changed so to preserve energy and
enstrophy, i.e. to mimic the two-dimensional Navier-Stokes equation
\cite{ABCFPV,DitMog,gilbert2002inverse}. In two dimensions an inverse
(i.e. toward the large scales) energy cascade accompanied by a direct
cascade of enstrophy takes place. However, in the shell model it has
been shown that the latter is expected to display a spectrum akin to
the enstrophy equipartition one \cite{DitMog}. Similarly to the
extension of hydrodynamic turbulence to noninteger dimensions
\cite{lvov2002quasigaussian, frisch2012decimation}, in shell models a
tuning of the coupling parameter of nonlinear interactions can be
associated to a continuous variation of the system dimensionality. An
interplay of equilibrium and cascades is observed at varying this
parameter \cite{gilbert2002inverse, tomray}. It is thus interesting to
inquire whether the equilibrium or non-equilibrium character can be
distinguished using the tools introduced in this work. Even more
interesting would be to adapt our tools to the case of
three-dimensional Navier-Stokes turbulence and, in particular, to
explore their behavior at scales larger than the forcing scale where
some features seem to be ascribable to equilibrium
\cite{dallas2015statistical,alexakis2019thermal,gorce2022statistical}
while others show deviations from equilibrium \cite{ding2023departure,
  hosking_schekochihin_2023}.  It could be not surprising, but surely
very interesting, to discover that some properties are well captured
by the equilibrium physics and others are not.

\begin{acknowledgements}
We are grateful to L. Biferale for useful discussions and to
S. Chibbaro for his feedback on the original draft. A.V. acknowledges
the support from the MIUR PRIN 2017 Project 201798CZLJ,
‘Coarse-grained description for non-equilibrium systems and transport
phenomena’ (CO-NEST).  M.C. acknowledges the Iniziativa Specifica
INFN-FIELDTURB.
\end{acknowledgements}

\appendix
\section{Numerical details of simulations\label{app:numerics}}

The Sabra model (\ref{eq:shell}-\ref{eq:sabra}) was integrated by
means of a 4th order Runge-Kutta algorithm, with exact integration of
the linear term  (see e.g. \cite{gallavotti}). In the
turbulent case, the shell velocities were first initialized according
to the scaling law: $|u_n| \sim k_n^{-1/2}$, and each complex velocity
with a random phase. In the inviscid and unforced shell model we
fixed the total energy $E=0.13$ and distributed it equally among all
shells as $|u_n| = \sqrt{2E/N}$, with phase assigned randomly as
well. Then, in both cases, we let the system evolve for a long enough transient of
time (many turnover times) until a stationary state is reached, after
which we start our measurements.  As for the forcing $f_n$ in
Eq.~(\ref{eq:shell}), we inject energy at scale $n_f$ and $n_f+1$ by
imposing
\begin{equation}
  f_k = \left\{
  \begin{array}{ll}
    \epsilon/(2u_k^*) &\mathrm{for}\quad k=n_f,n_f+1\\
    0                 &\mathrm{otherwise}
  \end{array}
  \right.
  \end{equation}
so that the energy power $\sum_n \Re(f_n u_n^*)=\epsilon$ is constant. In all our simulation $\epsilon=1$.

Table~\ref{tab:param} summarizes the parameters
we used in the simulations.
\begin{table}[h]
  \centering
  \begin{tabular}{|c|c|c|c|}
    \hline
    \textbf{Parameter} & \textbf{run-LSF} &  \textbf{run-EQ} & \textbf{run-ISF} \\
    \hline
    \rule{0pt}{2ex} $\Delta t$ & $5 \cdot 10^{-5}$ & $5 \cdot 10^{-5}$  & $10^{-5}$ \\
    \rule{0pt}{2ex} $N$ & 24 & 15  & 30\\
    \rule{0pt}{2ex} $k_0$ & $2^{-4}$ & $2^{-4}$ & $2^{-10}$\\
    \rule{0pt}{2ex} $n_f$ & 1 & N.A. & 13\\
    \rule{0pt}{2ex}$\nu$ & $10^{-6}$ & $0$  & $5\cdot 10^{-7}$\\
    \hline
  \end{tabular}
\caption{Values of the parameters used in numerical
  simulations. Run-LSF and Run-ISF correspond to the forced shell
  model, i.e. the turbulent case, with forcing at large scales or
  intermediate scales, respectively, while run-EQ denotes the
  equilibrium case, i.e. the unforced inviscid shell model.  The
  parameters are the time step ($\Delta t$), the number of shells ($N$), the
  smallest wave number ($k_0$), the shell number where the forcing is
  acting ($n_f$), and the viscosity ($\nu$). In the forced runs the
  forcing acts on shell $n_f$ and $n_f+1$ in such a way to fix the
  energy injection rate to $\epsilon=1$.\label{tab:param}}
\end{table}

\section{Issues on the numerical computation of
asymmetric correlation functions $\Psi_{e_n}(t)$ \label{app:Pomeau}}
In this Appendix we discuss some delicate issues about the numerical
computation of the correlation functions (\ref{eq:psi}) and its
equivalent form (\ref{eq:phi}).
First of all, we can rewrite Eq.~(\ref{eq:phi}) as
\begin{equation}
  \Phi_x(\tau)=  \Theta_x(\tau)+\Psi_x(\tau)\,,
\end{equation}
with $\Theta_x=\frac{1}{3}(\langle x^3(t+\tau)\rangle -\langle
x^3(t)\rangle)$.  Owing to the assumed stationarity of $x(t)$, it must
hold $\Theta_x(\tau)=0$, and thus $\Phi_x(\tau)\equiv\Psi_x(\tau)$.
However, in numerical evaluation, due to the way cancellations are
realized, the two equivalent functions $\Psi_x$ and $\Phi_x$ have
different pros and cons, as discussed below. In the following we will
drop the subscript $_x$ and the time dependence from the $x$ variable
to ease the notation.

\begin{figure}[b]
\centering
\includegraphics[width=.5\textwidth]{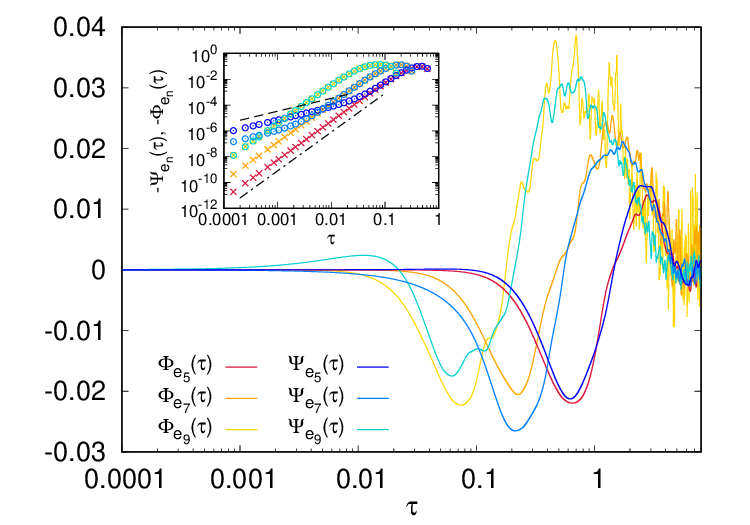}
\caption{ 
  Correlation function $\Psi_{e_n}(\tau)$ and $\Phi_{e_n}(\tau)$ for
  $n=5,7,9$ versus $\tau$. Statistics is over
  $5\cdot 10^5$ samples. Inset: short time behavior ($\tau<1$) of
  the same correlation functions with inverted sign and in logarithmic $y$ scale. 
  The dashed and dash-dotted lines denote respectively a linear and a cubic dependence on $\tau$. 
  Statistics over $10^6$ samples.
  Notice that computing the running average
  (not shown) of $\langle e_n^2\dot{e_n}\rangle= d/dt \langle
  e_n^3\rangle/3$ one can clearly see the slow convergence to zero,
  which is responsible for the spurious linear behavior.}
  \label{fig:pomeau-comp}
\end{figure}

The obvious advantage of computing $\Psi(\tau)$ is that, especially at
large $\tau$, it guarantees the cancellation of the term
$\Theta(\tau)$, as this is automatically imposed. This can be
appreciated from Fig.~\ref{fig:pomeau-comp} where one can see that,
at large $\tau$, the curves obtained computing $\Psi$ tend to be
smoother than those obtained by computing $\Phi$, although both curves
do convey the same result.
However, as it will be shown below, problems due to statistical
convergence can manifest at small $\tau$, while the computation of
$\Phi(\tau)$ is more efficient in ensuring the cancellations at small
$\tau$'s.  To realize such issue it is useful to
expand in series $\Theta(\tau)$, $\Psi(\tau)$ and $\Phi(\tau)$:
\begin{eqnarray}
  \Theta(\tau)&=& \phantom{-}\langle x^2 \dot{x}\rangle \tau+ \frac{1}{2} \langle [2x\dot{x}^2+x^2\ddot{x}]\rangle \tau^2 \nonumber\\
  &+&\frac{1}{6}\langle[2 \dot{x}^3+6x\dot{x}\ddot{x}+x^2\dddot{x}]\rangle\tau^3 +\ldots \label{eq:deltaseries}\\
  \Psi(\tau)&=&-\langle x^2 \dot{x}\rangle \tau-\frac{1}{2}\langle [2x\dot{x}^2+x^2\ddot{x}]\rangle \tau^2\nonumber\\
  &-&\frac{1}{6}\langle [6x\dot{x}\ddot{x}+x^2\dddot{x}]\rangle\tau^3 +\ldots\,. \label{eq:psiseries}\\
\Phi(\tau)&=& \frac{1}{3} \langle \dot{x}^3\rangle \tau^3+\ldots\,.
\end{eqnarray}
Since $\Phi=\Psi$ also $\Psi(\tau)$ should behave the same way at small $\tau$. In order to see this it is useful to rewrite Eqs.~(\ref{eq:deltaseries}-\ref{eq:psiseries}) as
\begin{eqnarray}
  \Theta(\tau)\! &=&\!  \frac{1}{3} \frac{d}{dt}\langle x^3\rangle \tau+ \frac{1}{2} \frac{d}{dt}\langle [x^2\dot{x}]\rangle \tau^2+\frac{1}{6} \frac{d^2}{dt^2}\langle x^2\dot{x}\rangle\tau^3  \phantom{\qquad}\label{eq:deltaseries2}\\
    \Psi(\tau)&=&-\frac{1}{3} \frac{d}{dt}\langle x^3\rangle \tau- \frac{1}{2} \frac{d}{dt}\langle [x^2\dot{x}]\rangle \tau^2\nonumber \\
    &-&\frac{1}{6}\left[ \frac{d^2}{dt^2}\langle x^2\dot{x}\rangle-2\langle \dot{x}^3\rangle\right]\tau^3 \,. \label{eq:psiseries2}
\end{eqnarray}  
Now, clearly all the terms of the form $d^{k}/dt^k \langle
[\ldots]\rangle$ should vanish by stationarity, so that
$\Psi(\tau)=\frac{1}{3} \langle \dot{x}^3\rangle \tau^3$. However, the
statistical convergence of $d^{k}/dt^k \langle [\ldots]\rangle \to 0$
may be hard to be obtained with a finite statistics leading to
spurious $O(\tau)$ terms. The inset of Figure~\ref{fig:pomeau-comp} does
illustrate precisely this problem. As one can see the small $\tau$
behavior of $\Psi$ when computed for the energy of shell $5$ and $7$
starts with a spurious linear behavior and does recover the correct
$\tau^{3}$ behavior only at sufficiently large $\tau$'s, while $\Phi$
does always display the correct $\tau^{3}$ dependence.  For shell $9$
the two curves do coincide, this is simply due to the fact that
shell $9$ is much faster than $7$ and $5$ so that statistical
convergence, and thus the cancellations, can be realized more easily.

We conclude this appendix by noticing that in Ref.\cite{Joss}, it was
reported a linear behavior at small $\tau$ for an experimental time
series of a turbulent velocity flow. As discussed above, we strongly
believe that the linear behavior have a spurious origin. It would thus
be very interesting to re-analyze the data using $\Phi(\tau)$ instead
of $\Psi(\tau)$ for confirmation.

\begin{figure}[b!]
    \centering
    \includegraphics[width=.5\textwidth]{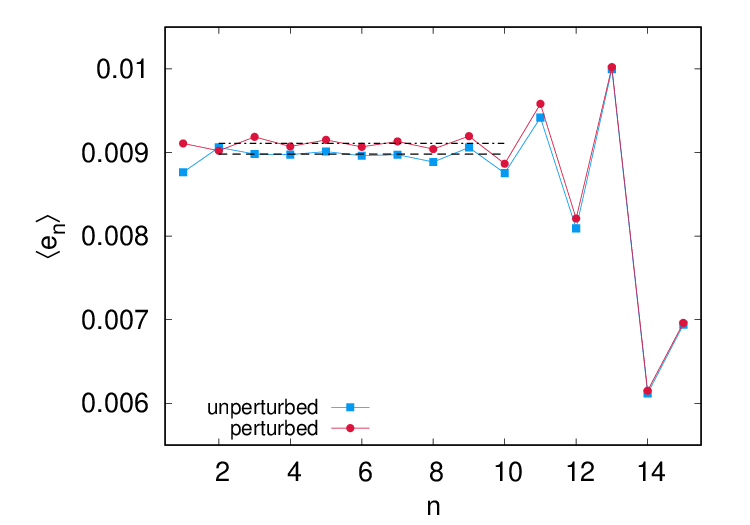}
    \caption{Energy spectra of unperturbed and perturbed inviscid
      Sabra shell model. The fit on the equiparted region of the
      spectra, appertaining to the shells which are far enough from
      the lower boundary, lead to the value
      $R^{(\infty)}_{corr}$. Initial perturbation $\delta_{m=7}\simeq
      1.79\cdot10^{-3}$.
    \label{bothspectra}}
\end{figure}

\section{Energy equipartition in the inviscid shell model \label{app:equilibrium}}
As discussed in App.~\ref{app:numerics}, for the inviscid shell model
we start from an initially equiparted spectrum. However, as shown in
Fig.~\ref{bothspectra}, after a long average the energy spectrum shows
clear departures from equipartition at large shell indexes, where
oscillations do appear. A few more simulations (not shown), performed
at changing the number of shells while keeping the energy per shell
constant in the initial condition, demonstrate that the oscillations
remain confined to the last 5-6 shells. Indeed by appropriately
shifting the shell axis we observed that the oscillations do
superimpose.  This demonstrates that such oscillations are due to the
boundary conditions which in the shell model, having a single variable
per shell, are expected to have a stronger effect than in the
truncated NSE. The main consequence of such oscillations is to alter
the equipartition value of the energy far from the boundary, so to
invalidate the naive expectation $R_{eq}^{(\infty)}=1/N$ for the
asymptotic value of the energy response functions, as discussed in the
main text. In order to determine the corrected (effective) value we
have run a long simulation after a perturbation and computed the
average spectrum: this way one can directly measure the energy shift
in the far from boundary shells, which display equipartition (see
Fig.~\ref{bothspectra}). Due to this effect, the actual difference
between the two equipartition values is larger than it would be in the
case of perfect equipartition, meaning that the asymptotic value of
$R_{m,n}(t)$ is slightly larger than expected (see
Figure~\ref{fig:Resp_Turb-Eq_E}(b)).

\section{Initial time-derivatives of $R_{m,n}(t)$ \label{appendixdRdt}}

In this Appendix we focus on the initial time derivatives of the
energy response functions, showing that they provide information on
the direction of the average energy flux among shells (where present).

From Eqs.~(\ref{eq:shell}-\ref{eq:sabra}) we can derive the following
equation for the rate of change of energy at shell $n$:
\begin{equation}
    \frac{de_n}{dt} = I_n + T_n + D_n\,,
    \label{enebalance}
\end{equation}
where the three terms are the input power $I_n = \Re\{f_n u_n^*\}$, the non-linear transfer of energy
\begin{equation}
    T_n = -[\Delta_{n+1}-\tfrac{1}{2}\Delta_n-\tfrac{1}{2}\Delta_{n-1}]\ ,
    \label{Tn}
\end{equation}
and the dissipated power $D_n= -\nu k_n^2 |u_n|^2=-\nu k_n^2 e_n$.
For the sake of readability we omitted the time dependence in all the
dynamical quantities.  The transfer term explicitly shows that, by
construction of the model, energy is directly exchanged between
neighbors and next-to-neighbors shells, i.e. within a ``range'' (or
distance) $2$.  The real quantities $\Delta_n$,  defined as
$\Delta_n = k_n \Im\{u^*_{n-1} u^*_{n} u_{n+1}\}$ (see also Eq.~\eqref{eq:flux}),
play a key role in the following argument.

In both inviscid and turbulent cases (assuming in the latter to
restrict ourselves in the inertial range so that the forcing is absent
and the dissipation can be neglected) we can express the initial time
derivative of $R_{m,n}$ using \eqref{enebalance} as follows:
\begin{equation}
    \frac{dR_{m,n}}{dt}\bigg\rvert_{t=0} =\frac{1}{\delta_m} \left[ \overline{T'_n(t)}-\overline{T_n(t)}\right]\bigg\rvert_{t=0}\ ,
    \label{dRdt}
\end{equation}
where primed quantities refer to the perturbed system.

Let us start from the inviscid case. As shown in the inset of
Fig.~\ref{fig:Resp_Turb-Eq_E}b, the time-derivative of $R_{m,n}$ is
zero for $t=0$ for any $n$, which can be understood as follows.
Clearly, if $|n-m|>2$ the perturbed shell $m$ does not directly
interact with shell $n$, which is thus unaware of the perturbation for
some time and, consequently, $\displaystyle
dR_{m,n}/dt\big\rvert_{t=0} = 0$. On the other hand if $|m-n|\leq 2$ the shell $n$ will in principle be affected by energy perturbation via the non-linear term.
However, at equilibrium,  we should expect
the shell velocities to be statistically
independent and Gaussian. Now, by using Eq.~(\ref{enebalance}) with $I_n=D_n=0$, for $n=m+1$ it is easy to derive that
\begin{eqnarray}
  \frac{dR_{m,m+1}}{dt}\bigg\rvert_{t=0} &\propto& \tfrac{1}{2} (\overline{\delta \Delta_{m+1}} + \overline{\delta\Delta_{m}}) \bigg\rvert_{t=0}\ .
\end{eqnarray}
where $\delta \Delta_m = k_m \Im\{u^*_{m-1} \delta u^*_{m} u_{m+1}\}$
and $\delta \Delta_{m+1} = k_{m+1} \Im\{\delta u^*_{m} u^*_{m+1}
u_{m+2}\}$, and in general $\delta f$ denotes the difference between
variable $f$ in the perturbed and unperturbed systems.  Under the
assumption that the average over many realization is equivalent to the
statistical average, one finds that the triple moments factorize into
the product of three single moments, two of which are zero.  With the
same reasoning one can conclude that the derivative should be zero
also for $n=m-2,m-1$ and $m+2$.

\begin{table}[b!]
    \centering
    \begin{tabular}{|c|c|}
        \hline
        {$n$} & \textbf{R.h.s. of \eqref{numdRdt}} \\
        \hline
        \hline
        \rule{0pt}{3ex}$m-2$ & $ -\overline{\alpha_m\Delta_{m-1}} \big\rvert_{t=0}<0$ \\[0.8ex]
        \hline
        \rule{0pt}{3ex}$m-1$ & $\overline{\alpha_m \left[ \tfrac{1}{2}\Delta_{m-1}-\Delta_{m} \right]}\big\rvert_{t=0}<0$ \\[0.8ex]
        \hline
        \rule{0pt}{3ex}$m+1$ & $\overline{\alpha_m \left[ \tfrac{1}{2}\Delta_{m} + \tfrac{1}{2}\Delta_{m+1} \right] }\big\rvert_{t=0}>0$ \\[0.8ex]
        \hline
        \rule{0pt}{3ex}$m+2$ & $\overline{\alpha_m \tfrac{1}{2}\Delta_{m+1}} \big\rvert_{t=0}>0$ \\[0.8ex]
        \hline
    \end{tabular}
    \caption{Initial time derivatives of $R_{m,n}(t)$ for $|n-m| \leq 2$.}
    \label{tabdRdt}
\end{table}

We now discuss the turbulent case. For $|m-n|>2$ the same reasoning
relying on the distance between perturbation and response applies, so that the initial derivatives of the RFs should be zero. For
$|m-n| \leq 2$, one can construct the following argument.
First, we rewrite the perturbed velocity \eqref{uprime} in terms of 
$u_m$ as $u'_m = \sqrt{1+{\delta_m}/{e_m}} u_m$, so that 
\begin{equation}
    \delta u_m \!=\! (u'_m\!-\!u_m)\!=\!\left( \sqrt{1+\frac{\delta_m}{e_m}} \!-\! 1 \right)u_m\!\equiv\!\alpha_m u_m\,,
    \label{deltau}
\end{equation}
where we notice that $\alpha_m$ is a real positive quantity for $\delta_m>0$.
Then, the equation \eqref{dRdt} can be explicitly written as:
\begin{eqnarray}
    \delta_m \frac{dR_{m,n}}{dt}\bigg\rvert_{t=0}\!\!\! =\! -\!\biggl[\overline{\delta\Delta_{n+1}}\!-\!\tfrac{1}{2}\overline{\delta\Delta_n}\!-\!\tfrac{1}{2}\overline{\delta\Delta_{n-1}}\biggr]\bigg\rvert_{t=0}\,,
    \label{numdRdt}
\end{eqnarray}
where 
\begin{eqnarray}
&&    \delta\Delta_{n}\big\rvert_{t=0} \!= \!k_n\Im\biggl[\delta_{m,n-1} \alpha_{n-1} u^*_{n-1} u^*_{n} u_{n+1}\!\label{deltaDeltaexplicit} \\ &&+\! \delta_{m,n}\ \alpha_n u^*_{n-1} u^*_{n} u_{n+1} + \delta_{m,n+1} \alpha_{n+1} u^*_{n-1} u^*_{n} u_{n+1}\biggr]
    \nonumber
\end{eqnarray}
where the Kronecker delta $\delta_{a,b}$ imposes that at least one of the three shell velocities involved has to be the initially-perturbed one, otherwise there will be no contribution, as seen in the previous case.

Computing the sign of the initial time derivative of the RFs amounts to studying terms of the kind (indices omitted):
\begin{equation}
    \overline{\delta\Delta} \sim k\overline{\alpha\Im\bigl\{uuu\bigr\}} = \overline{\alpha\Delta} ,
    \label{deltaDeltanoindices}
\end{equation}
but $\alpha$ is positive by definition (see Eq.~\eqref{deltau}), and it can be shown \cite{Sabra} that $\langle\Delta\rangle$ is positive on average, as it is related to the energy flux \eqref{eq:flux}. Again we will assume that $\overline{\Delta}=\langle\Delta\rangle$.

Once established this result let us reconsider \eqref{numdRdt} and
\eqref{deltaDeltaexplicit}. In Table \ref{tabdRdt} we explicitly write
the non-zero terms, when varying $n$ in the range of indices we are
studying.  Given that $\overline{\Delta_n} > 0$, at least for $n$ in
the inertial range, we have that backward RFs start with negative
slope, while forward ones with positive slope. Concerning the case
$n=m-1$, where the sign of the expression is not as straight-forward
as the others, the negativity of the r.h.s. comes from an explicit
result found in Eq. (16) of \cite{Sabra}.

\end{document}